\documentclass[sigconf,natbib=false,backend=biber,datamodel=acmdatamodel,style=acmnumeric]{acmart}
\usepackage{graphicx} 
\usepackage{caption}
\usepackage{balance}
\usepackage[backend=biber]{biblatex}

\begin{document}
\title{Integrating a Python Dynamical core into ICON}
\author{Mauro Bianco}
\author{Till Ehrengruber}
\author{Enrique González Paredes}  
\author{Andreas Jocksch}
\author{Christos Kotsalos}
\author{Ioannis Magkanaris}
\author{Philip Müller}
\author{Edoardo Paone}
\author{Mikael Simberg}
\author{Hannes Vogt}
\affiliation{%
  \institution{ETH Zurich, Swiss National Supercomputing Centre (CSCS)}
  \city{Lugano}
  \country{Switzerland}
}
\author{Jacopo Canton} 
\author{Yilu Chen}
\author{Anurag Dipankar}
\author{Nicoletta Farabullini}
\author{Michael Jähn}
\author{Matthieu Leclair}
\author{Ong Chia Rui} 
\affiliation{%
  \institution{ETH Zurich, Center for Climate System Modeling (C2SM)}
  \city{Zurich}
  \country{Switzerland}
}
\author{Nathan Beech}
\author{Nicolas Gruber}
\affiliation{%
  \institution{ETH Zurich, Department of Environmental Systems Science}
  \city{Zurich}
  \country{Switzerland}
}
\author{Christoph Müller}
\author{Daniel Hupp}
\author{Xavier Lapillonne}
\affiliation{%
  \institution{Swiss Federal Office for Meteorology and Climatology (MeteoSwiss)}
  \city{Zurich}
  \country{Switzerland}
}

\begin{abstract}
The transition of Earth-system models to exascale is often hindered by rigid, monolithic Fortran codebases and maintenance-heavy compiler directives. While high-level DSLs offer a solution, they frequently fail due to cumbersome integration. We present the integration of a Python-based ICON dynamical core into the original Fortran simulation code. Leveraging the GT4Py DSL and the Data-Centric (DaCe) optimization framework, we demonstrate that high-level Python can be seamlessly integrated into legacy infrastructure without performance loss.

Our results challenge the assumption that Python orchestration introduces prohibitive HPC overhead. In production-grade global simulations, our Python dynamical core achieves a 20--30\% performance improvement over the highly-optimized Fortran+OpenACC implementation, with a 10\% improvement on the total time for a coupled setup.
Driven by advanced data-flow optimizations and automated kernel fusion, this approach replaces hardware-entangled directives by generating optimized device code from a single, portable Python source.
This work proves that Python can provide a sustainable, efficient, and hardware-agnostic future for global climate modeling.

\end{abstract}



\maketitle

\section{Introduction}

Climate research has become one of the most consequential scientific domains, both for advancing our fundamental understanding of the Earth system and for addressing the societal challenges posed by climate change. For decades, progress in this field has been driven by sophisticated numerical models implemented in large, imperative codebases, often written in Fortran and developed over decades by many contributors. While immensely valuable, these legacy implementations have grown monolithic, with complex control flow, global state, and limited modularity. As a result, they are difficult to extend, test, and maintain.

Portability across modern computing architectures has traditionally been achieved by embedding compiler directives around computational kernels. Although effective at the time, this approach has gradually entangled the scientific code with architecture‐specific optimizations. The outcome is a proliferation of conditional compilation and platform‑dependent code paths that obscure the underlying algorithms and impede transparent testing, refactoring, and innovation.

This situation poses a significant challenge as computational science increasingly embraces data‑centric methodologies --- particularly machine learning, artificial intelligence, and hybrid modeling approaches. To support such emerging workflows, the computational tools used in climate science must be rethought and modernized. They need to be modular, maintainable, interoperable with new ecosystems, and flexible enough to incorporate novel methods without destabilizing established production workflows.

To address these needs, we utilize GT4Py, a Python embedded domain-specific language (DSL) that allows climate scientists to express numerical methods at a high level while targeting optimized backends for modern heterogeneous architectures. Using GT4Py, we implemented a Python‑based version of the ICON model. In this paper, we demonstrate how the Python implementation of ICON’s dynamical core can be seamlessly integrated into the original Fortran codebase. Our approach yields performance exceeding the existing OpenACC implementation, improves modularity and testability, and is capable of supporting kilometer‑scale production simulations of the Earth system.

This work serves as both a proof‑of‑concept and a validation framework for the Python implementation. More broadly, it challenges the persistent assumption that Python is unsuitable for high‑performance computing, showing instead that, when paired with modern compiler and DSL techniques, it can effectively support large‑scale Earth system modeling.

The paper is organized as follows: in Section \ref{sec:related} we highlight related work and then describe the baseline implementation from which we start in Section \ref{sec:baseline}. In Section \ref{sec:impl} we describe the details of the implementation and its novelty, while Section \ref{sec:results} will focus on results. Finally we conclude the paper with Section \ref{sec:conclusion}.

\section{Related Work}
\label{sec:related}
The literature on Earth system modeling is vast, but the efforts towards a fully portable model demonstrating exascale worthy simulations are only a handful. The Simple Cloud-Resolving E3SM Atmosphere Model (SCREAM; \cite{Donahueetal2024,wcd-6-447-2025,ubbiali-2025}) used the \textsc{C++} library Kokkos \cite{Trottetal2022} to port the entire atmospheric component of the Energy Exascale Earth System Model E3SM. They demonstrated a throughput of 26 Simulated Days Per Day (SDPD) for an atmosphere-land coupled configuration at a grid spacing of 1.25 km globally, when appropriately rescaled \cite{GB2025MPI-M}. The Nonhydrostatic Icosahedral Atmospheric Model (NICAM; \cite{satohetal2014}) demonstrated a throughput of 17 SDPD in a similar configuration when using the Fortran code run on the CPUs. 

The most related work to the one we are presenting is \cite{GB2025MPI-M} in terms of simulation complexity and the code base. The Authors present a fully atmosphere-land-ocean-carbon coupled simulation performed using the Fortran+OpenACC version of ICON, with the use of DaCe \cite{dace2019} to optimize the dynamical core, utilizing the DaCe Fortran frontend. The throughput ICON demonstrated is about 146 SDPD, which is considerably higher than any of the models reported earlier at 1.25~Km resolution. Direct comparison with this paper is not trivial since the configurations we test are not identical to those.

We also use DaCe for our optimization of GT4Py programs, which are written in Python and provide, in our opinion, an easier integration with the data structures used in DaCe (the Stateful Data-Flow Graphs---SDFGs). In addition our work re-implements and optimizes the whole dynamical core of ICON, including the numerical diffusion as shown in Figure~\ref{fig:dycore}. 

\section{Background and baseline}
\label{sec:baseline}
A numerical Earth‑system climate and weather model is composed of several interacting components. The primary subsystems are the atmosphere, the ocean, and the land surface. These components exchange information with one another, but they evolve at different characteristic timescales. For example, atmospheric dynamics changes rapidly, while ocean dynamics evolves much more slowly.

The coupling between land and atmosphere is particularly tight. Land‑surface temperature and humidity responds quickly to atmospheric conditions and the position of the sun. Because of this, land can be represented using a relatively shallow model that mainly exchanges heat with only the lowest atmospheric layers. Although the land component is somewhat unstructured, relying heavily on lookup tables describing position and time dependent land‑surface properties, it is not computationally intensive.

However, the tight coupling of land to the atmosphere makes data locality crucial. If the atmospheric model is executed on GPUs, the land component must also run on GPUs, not primarily for its performance, but to enable efficient data exchange between the two subsystems.

The ocean, in contrast, does not require frequent data exchange, and its dynamics is less computationally intensive. For this reason, it can run concurrently on CPUs, while the coupling can be done transferring the appropriate data between devices, but not at every timestep.

A full Earth-system simulation can then be tailored to run effectively on hybrid systems, as has been shown in \cite{GB2025MPI-M}.

In this paper we will focus our attention to the atmospheric part of the atmosphere-land-ocean coupled simulation, which is the most computationally demanding and also most complex portion of the Earth system to simulate. The atmospheric model itself comprises several components. The most computationally intensive and numerically complex one is the \textit{Dynamical Core}, which solves the friction-free Navier-Stokes equations for the three wind components, temperature, pressure, and various moisture species constituting the state of the atmosphere at a given time. The rest of the components use the state computed by the dynamical core to add to it the effects of atmospheric processes, like phase changes within a cloud (\textit{microphysics}), radiative effects due to solar and terrestrial waves, turbulence, etc. As a rule of thumb: the coarser the resolution of the simulation, the larger the number of additional sub-grid parameterizations that are necessary to account for such phenomena. For this reason science is always aiming at increasing the spatial and temporal resolution of the simulations, so that they can better approximate first-principle solutions of the equations. 

The simulation software we are using in this paper is ICON \cite{icon15}, specifically, the starting point for this work is the architecture described by Dipankar et al. \cite{gmd-19-713-2026}. In that paper, the authors presented an implementation of ICON in which the computational kernels of the ICON dynamical core are substituted with pre-compiled functions generated from GT4Py \cite{paredes2023gt4py,ubbiali-2025,dahm-2023}.

Building upon the foundation in \cite{gmd-19-713-2026}, we show an approach that takes model development closer to a composable and testable software system, and also breaks some perceived barriers that have hindered the development of more modern approaches for decades. Namely, the typical DSL-approach would be to write some specific components of the software using a domain specific language, then generate the optimized code with it, and link the optimized code to the application. This approach has several limitations, that explain why, albeit in principle a DSL could improve both productivity and performance, this approach has never gained enough momentum and received only limited adoption. Our approach is substantially different: the whole dynamical core of ICON has been implemented in Python, using GT4Py, and the Fortran application directly invokes the Python interpreter, which in turn executes the application, either using just-in-time compilation or ahead-of-time compilation, depending on the specific use case. 

This is an example of modular development: a component (in this case the dynamical core of ICON), could be implemented in different languages and approaches, as long as the interface is accessible. These components can use distinct testing methodologies, thus making the development of large scale complex software leaner and flexible. In fact, the Python/GT4Py implementation of the dynamical core uses other libraries, such as \texttt{NumPy}~\cite{harris2020array} and \texttt{CuPy}~\cite{cupy_learningsys2017} for array representation and has experimental support for JAX, including JAX's just-in-time compilation~\cite{jax2018github} which offers a path to integration with machine learning workflows. Additionally, GT4Py uses DaCe~\cite{dace2019} for data-flow optimizations (see Section \ref{sssec:backend}) and GHEX \cite{ghex} to perform halo-updates. The latter point is especially important: the communication library in HPC applications, usually MPI, is usually considered a global choice for the whole application. GHEX can employ other transport layers to perform the actual halo-exchanges, like UCX or NCCL. These transport layers require an additional local synchronization to guarantee message ordering, but this happens only when entering and exiting the dynamical core. GHEX requires MPI only during initialization to map the ranks of the computation, but can be configured to use any transport layer to perform the actual communication.

\section{Implementation description}
\label{sec:impl}


\subsection{ICON4Py dynamical core}

\begin{figure}
    \centering
    \includegraphics[width=0.9\linewidth]{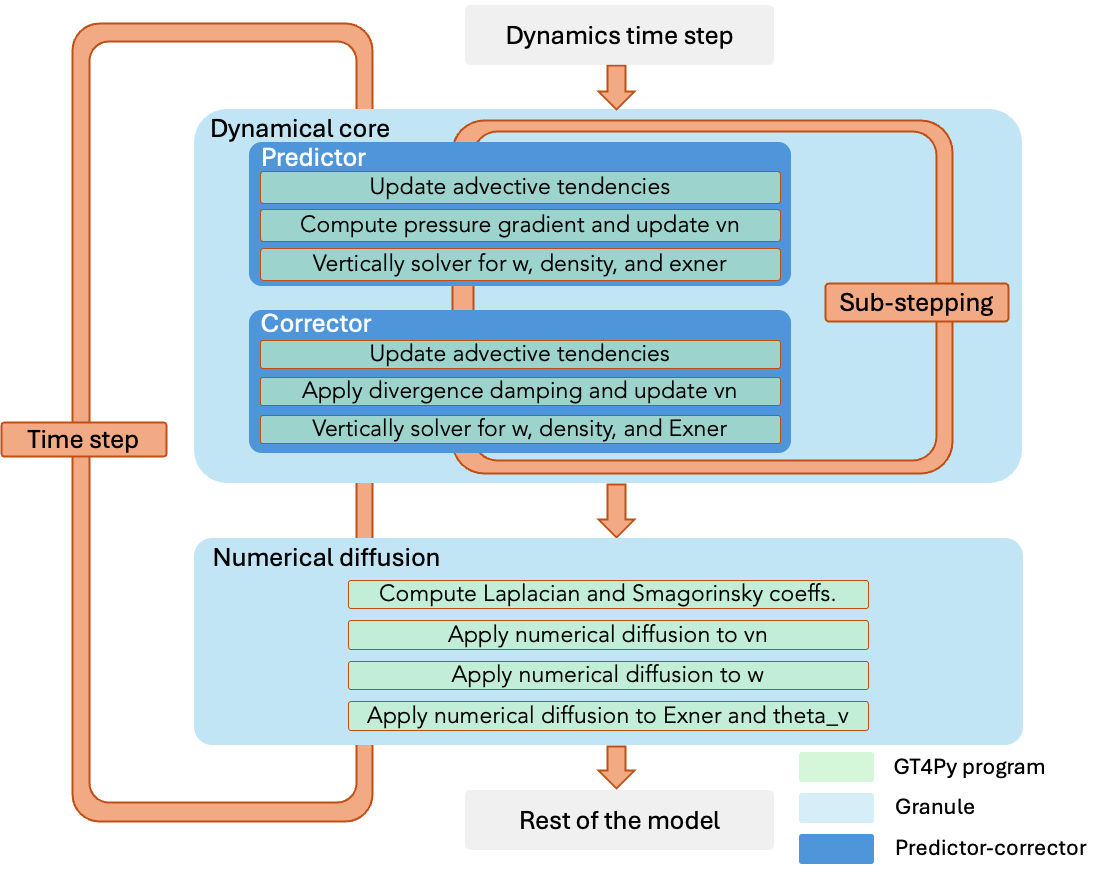}
    \caption{Schematics of the dynamics and its computation flow. The dynamics contains two parts: dynamical core, which solves the primitive equations, and numerical diffusion which applies diffusion to stabilize the simulations. The dynamical core performs a predictor and corrector steps in a finer time step than the main application, followed by a diffusion step. The portion indicated as "Dynamical core" is included in a Fortran routine called \texttt{nh\_solve}, while the "Diffusion" is in a routine called \texttt{nh\_diff}.}
    \Description[Schematics of the Dynamical core.]{Schematics of the Dynamical core computation flow.}
    \label{fig:dycore}
\end{figure}

The use of GT4Py is not limited to ICON. It has been used to write atmospheric components  of PMAP \cite{ubbiali-2025} and PACE \cite{dahm-2023}. To signify this difference, the re-write of ICON using GT4Py is termed as ICON4Py, and therefore the ICON4Py dynamical core refers to the ICON dynamical core written in GT4Py. It reflects the original ICON numerical discretization \cite{gmd-6-735-2013} and implements a nested time step and dynamical substep scheme for temporal integration.
The dynamical substep consists of a predictor-corrector scheme for the state variables.
Each substep begins with the computation of the advective tendencies.
The dynamical core then updates the horizontal momentum equations with the pressure-gradient and advection terms, applies divergence damping only in the corrector step, and advances the vertical momentum and thermodynamic state with a vertically implicit solver for numerical stability.

Diffusion is implemented as a separate \textit{granule} ("granule" is the term used in the ICON community to identify a software component to avoid confusion with the same term used in other contexts) that is called once per time step.  It supports a Smagorinsky plus fourth-order background formulation.
A diffusion step applies operators to horizontal wind, vertical wind, and optionally temperature and Exner fields, including lateral boundary and nudging treatments and increased coefficients for strong near-surface temperature gradients.

The dynamical core executes pre-compiled ICON4Py stencils over horizontal (triangular mesh) and vertical (regular) index ranges, and performs halo exchanges between subdomains with the GHEX \cite{ghex} library. As described above, GHEX is a library providing halo-exchange operations with several backends and allows expressing computation and communication overlap.

Compared to the original Fortran implementation, the Python framework allows for a clearer distinction between the individual components, e.g., between predictor and corrector. This makes the code more readable.
Another advantage of the current implementation is the thorough unit- and data-test infrastructure built around individual components which guarantees correctness of the code and allows for a more confident implementation of new features or modification of existing ones. This is applied both at the stencil level as well as at the granule level.

\subsection{GT4Py}
\label{ssec:gt4py}

 In this section we introduce GT4Py, a Python library implementing an embedded DSL for climate and weather numerical schemes which is central to the implementation of ICON4Py.

\subsubsection{GT4Py Concepts and Front-end}
\label{sssec:frontend}

\begin{figure}
    \centering
    \includegraphics[width=0.9\linewidth]{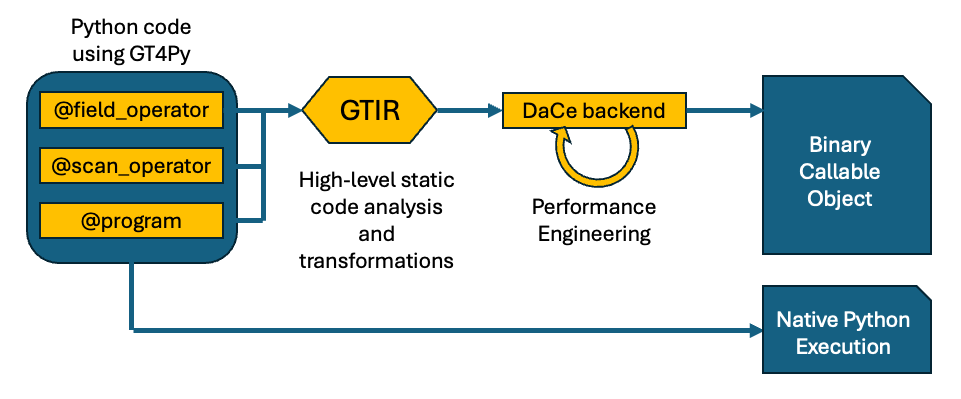}
    \caption{Depiction of the GT4Py workflow. A computation in GT4Py can be executed natively in Python or be processed by a transformation toolchain to produce efficient code on different architectures.}
    \Description[GT4Py workflow]{Depiction of the GT4Py workflow. A computation in GT4Py can be executed natively in Python or be processed by a transformation toolchain to produce efficient code on different architectures.}
    \label{fig:gt4py_workflow}
\end{figure}

\begin{figure}
    \centering
    \includegraphics[width=0.9\linewidth]{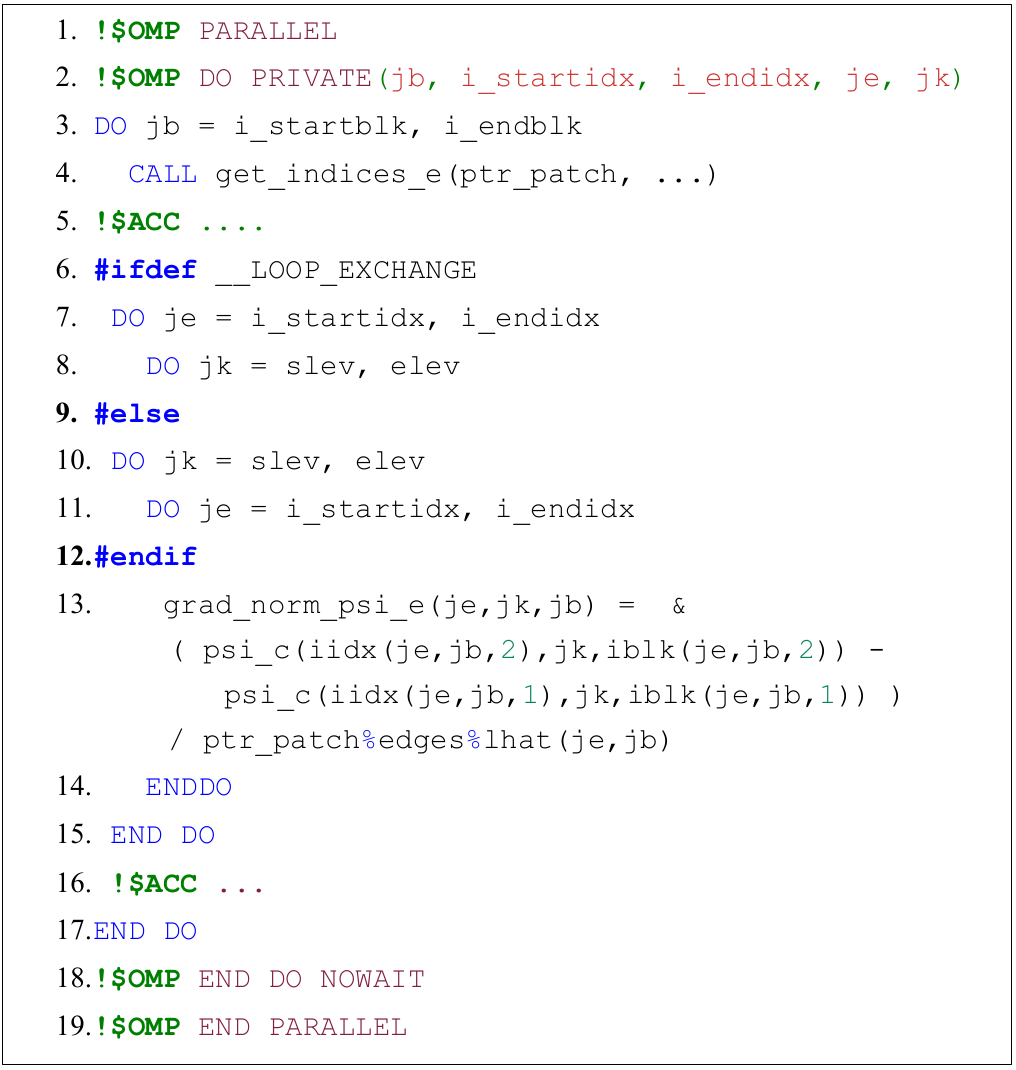}
    \caption{Sample code from Fortran implementation of ICON showing multiple compiler directives for optimization and compilation guards for customising loop order.}
    \Description[Sample code from Fortran implementation of ICON.]{Sample code from Fortran implementation of ICON showing multiple compiler directives for optimization and compilation guards for customizing loop order.}
    \label{fig:oacc_code}
\end{figure}

\begin{figure}
    \centering
    \includegraphics[width=0.9\linewidth]{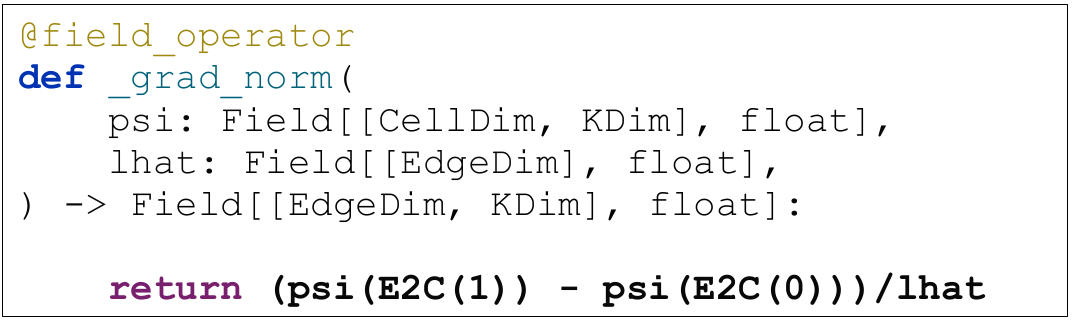}
    \caption{Sample code from GT4Py implementation of a numerical computation (taking values from the neighbor cells of an edge and computing on the edge).}
    \Description[Sample code from GT4Py]{Sample code from GT4Py implementation of a numerical computation (taking values from the neighbors cells of an edge and computing on the edge).}
    \label{fig:gt4py_code}
\end{figure}

The GT4Py Domain Specific Language is embedded in Python,
which allows GT4Py compute-intensive functions to coexist with
regular Python application logic. One important feature of GT4Py is to be able to run natively in Python, thus allowing debugging and seamless integration with other packages. The GT4Py semantics is then compatible with the Python one, even when we optimize the code to run efficiently on a variety of architectures and at scale. When enabling the performance optimized execution, the user does not need to change any code, while the integration with other libraries and tools remains untouched. 

GT4Py treats Python as a high-level front-end for defining numerical computations. This approach allows domain scientists to write atmospheric equations in a declarative, readable syntax while delegating hardware-specific optimizations to a specialized compilation stack. GT4Py provides a set of function decorators to the application developer to express some parallel compute patterns which are common in climate and weather applications, such as \texttt{@field\_operator}'s and \texttt{@scan\_operator}'s \cite{paredes2023gt4py}. A \texttt{@field\_operator} is a generalized stencil computation, while a \texttt{@scan\_operator} specifies operations that scan up or down the atmosphere. These two cover most of the computation in climate codes, not only dynamics, but also sub-grid parameterizations. The entry-points to the GT4Py DSL are annotated with the \texttt{@program} decorator. In Figure~\ref{fig:oacc_code} and Figure~\ref{fig:gt4py_code} we compare qualitatively a sample code found in the Fortran source code of the Fortran application (which contains different types of compiler directives and loop ordering guarded by compilation guards), and GT4Py code, which focuses on the numerical specification, leaving the optimization to the backend, providing a clear separation of concerns.

The GT4Py programs are taken by the GT4Py frontend to produce an internal representation of the computation, called GTIR. This is done completely in Python, without requiring any other external tool, like an external parser or command line tools to produce executable binaries. In the case of ICON, or in other grids defined by lookup-tables, computations involving neighbor accesses require an index lookup instead of a dot product of a multidimensional index with the grid strides (hence the term "stencil-like" and not "stencils"). GT4Py abstracts the way neighbor accesses are expressed, allowing similar syntax for structured (e.g. Cartesian) and unstructured grids. A GT4Py code, before the computations are specified, defines the fields types and their neighbor access patterns, so that GT4Py only needs to pass that information to the backend that will deal with code transformation and compilation for performance (see Figure \ref{fig:gt4py_code} for an example).

GT4Py uses a functional specification of the computations. The functional approach allows for natural function composition, which is used for both composing the computations and analyzing the access patterns. A \texttt{@field\_operator}, for instance, is written so it returns new fields with the result of the computation. This however does not imply that the fields are created and returned in the actual implementation. In fact, in order to avoid memory management problems and better utilize the memory system of the computing nodes, a \texttt{@program} specifies where the inputs and the outputs actually are in memory, breaking the functional composition when the binding is set. This allows the optimization passes to create intermediate storage if they are deemed necessary, but also to control the logic of the application while enabling performance optimization.

\subsubsection{GT4Py Backend}
\label{sssec:backend}
GT4Py code, written with the intention to enable developers to specify high-level abstract operations, must be transformed into an efficient implementation for the target architecture. This happens in two phases. A first phase is the high-level transformation of the functional specification into a format that highlights the data-flow requirements. This is done by applying inlining and substitutions on the GTIR to simplify the computations. For instance, the stencil-like computations can be coalesced to modify the tradeoff between computation and memory accesses.
Once this step is done, the computation is transformed to Stateful DataFlow Multigraphs (SDFGs) to be manipulated and optimized using the Data-Centric (DaCe) optimization backend \cite{dace2019}. This representation focuses on the movement of data rather than the reorganization of the computation (for instance, common sub-expression elimination). By analyzing the dataflow, the backend can perform sophisticated optimizations that are often difficult for standard Fortran compilers to automate. These are:

\begin{itemize}
\item Global Kernel Fusion: Reducing expensive store and loads from off-chip memory by merging multiple computational kernels into a single GPU operation.
\item Memory Layout Transformation: Automatically adjusting how arrays are stored in different layers of the memory hierarchy to maximize hardware throughput on different architectures.
\item Iteration Range Splitting: Decomposing the iteration space of a kernel into multiple sub-kernels to enable their integration with other kernels.
\item Use of specific hardware memory subsystems (caches, constant-memories, registers, etc.).
\end{itemize}

This toolchain allows us to address Portability of Performance: The same high-level Python code can be compiled and optimized for a variety of different hardware platforms, for instance NVIDIA GPUs, AMD GPUs, or for multi-core CPUs. The dynamical core remains "pure Python" from the developer's perspective, but the binary executed on the supercomputer is a highly tuned, hardware-specific artifact that rivals or exceeds the performance of hand-written code in imperative programming languages, such as, in our case, Fortran+OpenACC.

\subsection{Integration of ICON4Py into ICON Fortran}
\subsubsection{Interfacing Fortran and Python: py2fgen}
For convenient integration of the Python implementation into the Fortran-based ICON implementation, a small, feature-restricted tool \texttt{py2fgen}, was developed on CFFI \cite{cffi} which provides the embedding of the Python interpreter into the Fortran bindings library via ISO-C bindings. This translation layer is necessary because Fortran and Python (e.g., NumPy) utilize different shape descriptions formats for multi-dimensional arrays.

On the Python-side, the user decorates a function to be exported which serves two purposes: a) At bindings generation time, the C and Fortran types are extracted, either from type annotations or by explicitly providing a mapping from parameter name to type. Supported types are scalars with value semantics, and arrays with reference semantics (that is, arrays are not copied from Fortran to Python\footnote{With the exception of boolean arrays: due to the default size of 4 bytes for Fortran logicals we cannot trivially interpret the Fortran pointer with C/Python bool semantics. In our setup all boolean arrays are passed in the setup phase and are immutable, therefore we can copy the 4-byte-boolean arrays into 1-byte-boolean NumPy/CuPy arrays.}). b) At runtime raw Fortran/C  array descriptors consisting of pointer, data type and shape are translated into Python objects. Natively supported are transformations of array descriptors to NumPy or CuPy arrays. Additionally, the user can provide transformations to custom Python objects, e.g., in our case, GT4Py fields. These transformations do not require data copies: only the organization of the shape and access information is adapted to the target data types. The results of the transformations are cached based on the array descriptor to avoid any extra Python overhead on the second call of the function with the same Fortran arrays. As the relevant ICON arrays are all statically allocated, we use a cache of size 2 for double-buffered arrays to keep all field objects on the Python-side cached for the runtime of the ICON application.

On the Fortran-side, we provide a generated Fortran subroutine which hides low-level details of passing Fortran arrays like extracting pointers and shape, including handling of OpenACC device arrays and optional (non-associated) arrays. Derived types are not supported by the generator which requires the user to explicitly unwrap these types into plain Fortran arrays. This is mainly due to the lack of shared standard layout specification for derived data types. At runtime, at the first call to one of the exported functions, the Python interpreter is spun up and kept alive by the CFFI-embedding.

\subsubsection{Modifications in ICON Fortran}

The ICON4Py dynamical core implementation can be used in the ICON model implementation by specifying the appropriate building flags and following the documentation. The modifications in the ICON model implementation that allow the use of the ICON4Py dynamical core are minimal. What is needed to make the adaptation is to gather the pointers to the fields needed by the dynamics and pass them to the C function that implements the interface to the dynamics. The modifications are then almost all contained in the section of the code interfacing with ICON4Py and are guarded by compilation guards. Few changes needed to be propagated in other parts of the code because of the use of Fortran datatypes, which can be difficult to interact with. 

The C-ABI compatible function to interface to ICON4Py does not depend on GT4Py details, since the code calls the Python interpreter directly, and thus all the difficult interfaces to the dynamic libraries generated by the toolchain are hidden from the Fortran code. This is different from the typical DSL implementation, in which the host application needs to know the conventions used by the DSL in order to use it. In this way the intrusion in the host applications does not need to be retouched every time the conventions in the underlying tools change, which is a curse of other domain specific languages implementations.

\subsubsection{Hiding Python Interpreter Overheads}

\begin{figure*}[t]
    \centering
    \includegraphics[width=\textwidth]{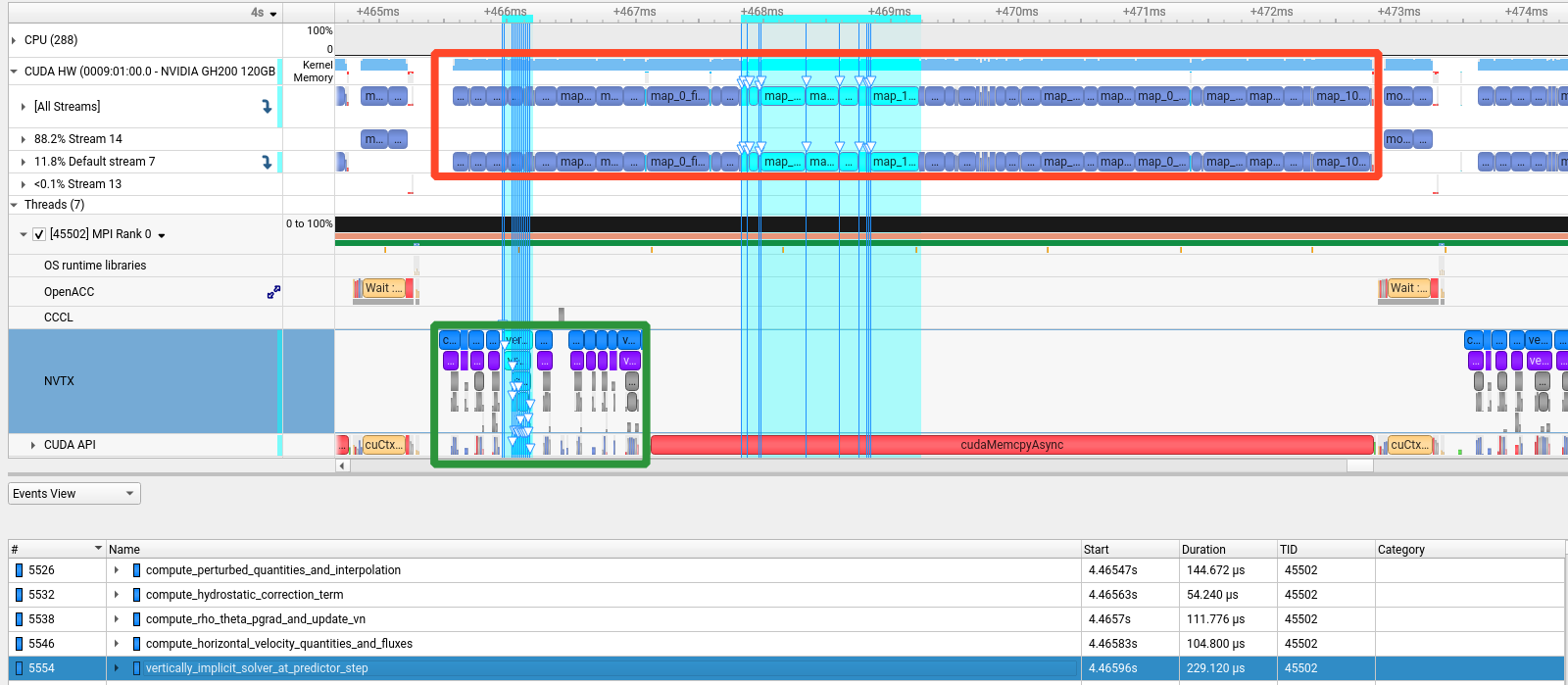}
    \caption{Execution profile of a substep in the ICON4Py dynamical core on a single GPU. The green box marks the time spent in the Python interpreter by GT4Py launching the GPU kernels of the substep, while the red box marks the time spent on the GPU executing the scheduled kernels. The cyan areas relate the scheduling time (first block) of the selected \texttt{vertically\_implicit\_solver\_at\_predictor\_step} to its execution time (second block), which is noticeably larger.}
    \Description[Execution profile of the ICON4Py dynamical core]{Execution profile of a substep in the ICON4Py dynamical core on a single GPU. The green box marks the time spent in the Python interpreter by GT4Py launching the GPU kernels of the substep, while the red box marks the time spent on the GPU executing the scheduled kernels. The cyan areas relate the scheduling time (first block) of the selected \texttt{vertically\_implicit\_solver\_at\_predictor\_step} to its execution time (second block), which is noticeably larger.}
    \label{fig:prof}
\end{figure*}

Because Python execution carries significant overhead, at simulation time Python is only used to dispatch the computationally intensive tasks to the compute devices. As an example, the execution profile of the first substep of the dynamical core, in Figure~\ref{fig:prof}, shows that the Python interpreter launches all the GPU kernels of the substep in a fraction of the time required for the actual computation work. Since the kernel scheduling operation does not wait until the actual computation on device has finished, the Python interpreter can continue launching kernels while the GPU is busy with computing, effectively overlapping kernel launching with computations already running on the accelerator. While the lead time before the first kernel starts running is undoubtedly present, it is minimal, and the advantage in performance, productivity and composability of the model code, in our opinion, more than compensates for it, providing better performance than traditional code while modernizing it substantially.

\subsection{Verification}

To ensure correctness, we build a verification hierarchy which, given the same input fields, checks the output fields of a code segment for a correctness requirement to be defined for each hierarchy level. Note that bit-wise identical computations are very hard to achieve since Fortran does not follow the IEEE 754 standard and allows for compiler optimizations assuming associativity of floating point computations, which our generated C++/CUDA code does not allow.

On the first level, we verify the correctness of segments of the dynamical core and diffusion that span between halo exchanges; the measure of error here is the relative error, with a standard tolerance of $10^{-12}$, which needs to be softened for certain stencils to $10^{-7}$ because of floating point issues such as cancellation mostly due to potential reordering of the operations that Fortran allows.
To implement this first level we introduce serialization in the ICON Fortran code and serialize the input and output fields of the code segment for specific experiment and namelist choice. We then run the corresponding code segment in Python with the serialized input fields and compare the output fields generated in Python with the serialized output fields.

On the second level of the verification hierarchy, we run the full ICON4Py dycore or diffusion, including the halo exchanges, by calling it in the Fortran host code and for comparison also run the original Fortran OpenACC code with the same input fields. In this mode copies of all input fields are generated which are fed to the ICON4Py versions of the dycore and diffusion, the Fortran version runs with the original fields, which ensures that the model continues with the fields of the original model not to generate consequential errors.
Again, all output fields are checked for relative errors similarly to the first verification hierarchy level. Contrary to the first level of the verification hierarchy, the second level does not utilize serialization at all, the verification happens as the Fortran ICON model is running.

The third and last level in the context of our work is \textit{probtest}, which is a numerical error growth test over 5-10 ICON timesteps which is equivalent to 25-50 dycore steps, assuming 5 dycore sub-steps per timestep.
For probtest one needs to generate a reference ensemble by running the same experiment multiple times with randomly perturbed inputs each run with a different RNG seed.
Usually, this reference ensemble is computed on CPU with the dynamical core and diffusion of the original Fortran code, and it is then used to generate reference values and tolerances for all of the output fields.
An ICON4Py GPU run can then be compared to these references and is declared passing if the errors are within the tolerances. Probtest was developed at MeteoSwiss and DWD, for more information, see \cite{icon_gpu} and \cite{probtest}.

\section{Results}
\label{sec:results}

\begin{figure*}[!ht]
    \centering
    \includegraphics[width=\textwidth]{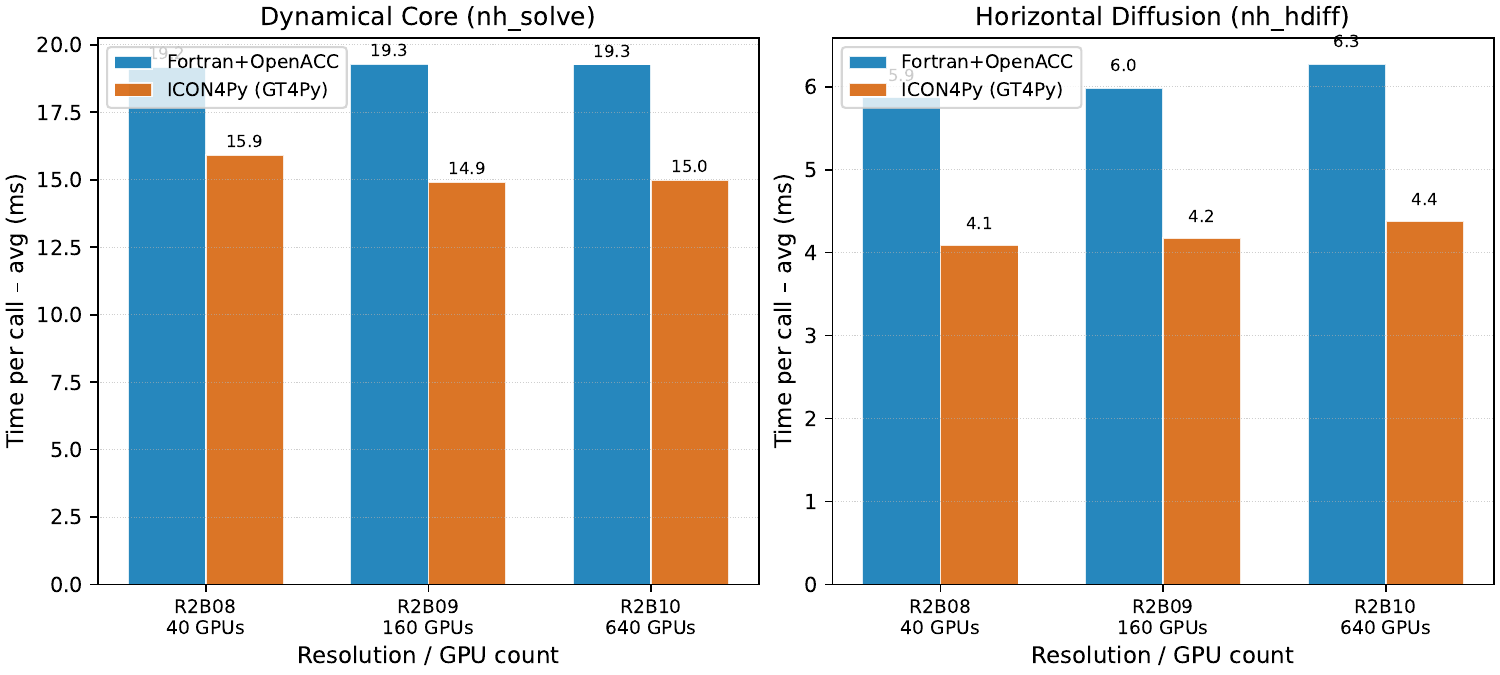}
    \caption{Weak scaling of the ICON4Py dynamical core compared to the OpenACC baseline across three ICON grid resolutions: R2B08 (40~GPUs), R2B09 (160~GPUs), and R2B10
    (640~GPUs). Each refinement level quadruples the horizontal cell count with a
    corresponding 4$\times$ increase in GPU count. Ideal weak
    scaling corresponds to constant bar height across resolutions. The left panel shows the dynamic substep \texttt{nh\_solve}, the right panel shows horizontal diffusion \texttt{nh\_hdiff}.}
    \Description[Weak scaling of dynamical core]{Weak scaling of \texttt{nh\_solve} (left) and \texttt{nh\_hdiff} (right)
    across three ICON grid resolutions: R2B08 (40~GPUs), R2B09 (160~GPUs), and R2B10
    (640~GPUs). Each refinement level quadruples the horizontal cell count with a
    corresponding 4$\times$ increase in GPU count. Bars show the average time per call
    for the Fortran+OpenACC baseline and the ICON4Py (GT4Py) implementation. Ideal weak
    scaling corresponds to constant bar height across resolutions.}
    \label{fig:weak_scaling_avg}
\end{figure*}

\subsection{Experimental setup}
\label{ssec:expsetup}

All benchmarks were executed on a Tier-0 supercomputer. Each compute node is equipped with four NVIDIA Grace Hopper Superchips, dubbed in the following as GH200 (system name hidden for double-blind review process).
Each GH200 node comprises 4 modules, each consisting of a Grace ARM CPU and a Hopper GPU, with 128GB of LPDDR RAM and 96 GB of HBM3 memory. The 4 modules are connected through NVIDIA NVLINK. 
The compute nodes are connected by an HPC Cray Slingshot-11 network with 200 Gbps injection bandwidth per module/GPU.

The benchmark results on the GH200 nodes are obtained by running the ICON Fortran application with the original Fortran OpenACC and our Python implementation. The times presented are the execution times of the dynamical core. The dynamical core represents, typically, 50\% of the overall application runtime, so the overall application benefit is approximately half of the one of the dynamics. We chose not to show the overall time, but only the one of the dynamical core for two reasons: first the dynamics and its implementation in Python is the actual focus of the paper, second, the overall time in actual simulations heavily depends on the configuration of the simulation itself. For instance, a scientist may increase the number of sub-steps, or increase the frequency of calls for other components, like radiation,affecting the impact of the performance improvements in the dynamics. The actual impact on the runtime is then dependent on these choices. We believe that the relevance of the paper is not diminished by this aspect, since our objective is to show that it is possible to engineer scientific applications to allow, not only high-performance, but also modularity and the adoption of new programming styles in HPC applications.


As depicted in Figure~\ref{fig:dycore}, the dynamics time step in ICON is split in two parts: the dynamical sub-stepping, and a diffusion step. Diffusion is part of the time step but it is not executed at every sub-step. The sub-step computation is enclosed in a Fortran routine called \texttt{nh\_solve}, while the diffusion is called \texttt{nh\_diff} (\texttt{nh} stands for \textit{non-hydrostatic}). We measure the performance using the ICON timing facilities, and make sure proper synchronization is performed. In this way we can fairly compare the two implementations and not hide overheads that can lay in the interface. As the \texttt{nh\_solve} routine runs a single sub-step it will be clear that the diffusion component is not really impacting the performance significantly.

ICON uses icosahedral grids.
A ``root'' subdivision (performed with the algorithm of Sadourny \cite{sadourny_integration_1968}) divides each of the 20 faces into $n^2$ equilateral triangles.
A second step recursively subdivides each triangle $m$ times into four smaller triangles by bisecting the edges.
The resulting grid is indicated as $RnBm$.
The resolution of the grid is the length of the final segment, after the last division. In our experiments, we split each edge of the icosahedron in 2 and we bisect the resulting triangles recursively either 8, 9 or 10 times, producing grids indicated with R2B08, R2B09 and R2B10, which corresponds to resolutions of 10~Km, 5~Km and 2.5~Km, respectively. The simulations we show are relevant examples of current and future production simulations both in terms of resolution and setup.



\subsection{Atmosphere-Land simulations}
\label{ssec:benchmarks}

In this Section we describe the execution times of atmosphere-land coupled simulations, as representative of a typical weather model, comparing the original OpenACC implementation and the ICON4Py dynamical core. The model configuration includes ICON's numerical weather prediction atmospheric (ICON NWP) and land components as described in \cite{Preinetal2026}. All the simulations use 120 vertical levels and a time step of 22 seconds. In Figure~\ref{fig:weak_scaling_avg} we show a weak scaling experiment. In order to do so we run larger and larger grids maintaining the same number of grid points per computing rank. As the amount of computation increases  four times when resolution is halved (keeping the same time step), we show how an R2B08 grid (10~Km resolution) on 40 GPUs compares with R2B09 (5~Km resolution) on 160 GPUs, and with R2B10 (2.5~Km resolution) on 640 GPUs. As we can see, the dynamical core (\texttt{nh\_solve}) and diffusion (\texttt{nh\_diff}) scale well, maintaining a constant load per GPU.

\begin{figure*}[tbp]
    \centering
    \includegraphics[width=\textwidth]{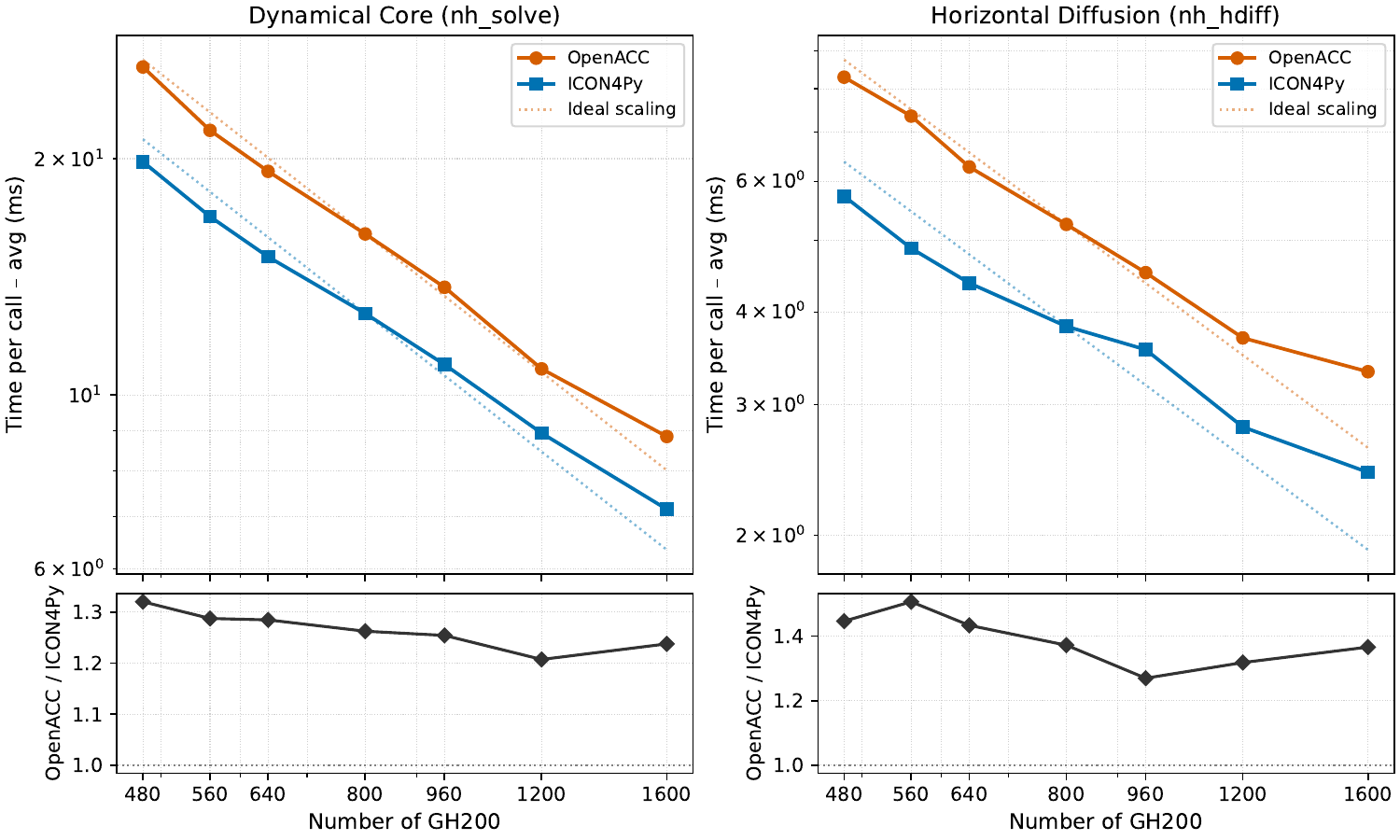}
    \caption{Strong scaling of the ICON4Py dynamical core compared to the
    OpenACC baseline on R2B10 (2.5\,km, 120 vertical levels) for atmosphere-land simulation. Top panels show the average wall time per call for
    the dynamic substep \texttt{nh\_solve} (left) and horizontal diffusion \texttt{nh\_hdiff} (right). Bottom panels show the
    performance ratio (OpenACC\,/\,ICON4Py), where values above~1 indicate ICON4Py is
    faster.}
    \Description[Strong scaling of the ICON4Py]{Strong scaling of the ICON4Py dynamical core compared to the
    OpenACC baseline on R2B10 (2.5\,km, 120 vertical levels) for atmosphere-land simulation. Top panels show the average wall time per call for
    the dynamic substep \texttt{nh\_solve} (left) and horizontal diffusion \texttt{nh\_hdiff} (right). Bottom panels show the
    performance ratio (OpenACC\,/\,ICON4Py), where values above~1 indicate ICON4Py is
    faster.}
    \label{fig:strong_scaling_R2B10}
\end{figure*}

In Figure~\ref{fig:strong_scaling_R2B10} we show strong scaling for the simulation from 480 to 1600 GPUs, for both \texttt{nh\_solve} and \texttt{nh\_diff}. In this more realistic example (the weak scaling, by keeping the time step constant, leads to non realistic simulations), we demonstrate how the performance of ICON4Py, as in all our benchmarks, measured from the Fortran side, shows 20--30\% improvement in dynamics, and up to 50\% for diffusion. The overall speed up of the simulation is 10--15\%, since the dynamics, in this simulation takes half of the computation time, and diffusion is only called at every time step and is less computationally demanding than \texttt{nh\_solve}.

\subsection{Atmosphere-Land-Ocean simulations}
\label{ssec:production}

\begin{figure}[tbp]
    \centering
    \includegraphics[width=\columnwidth]{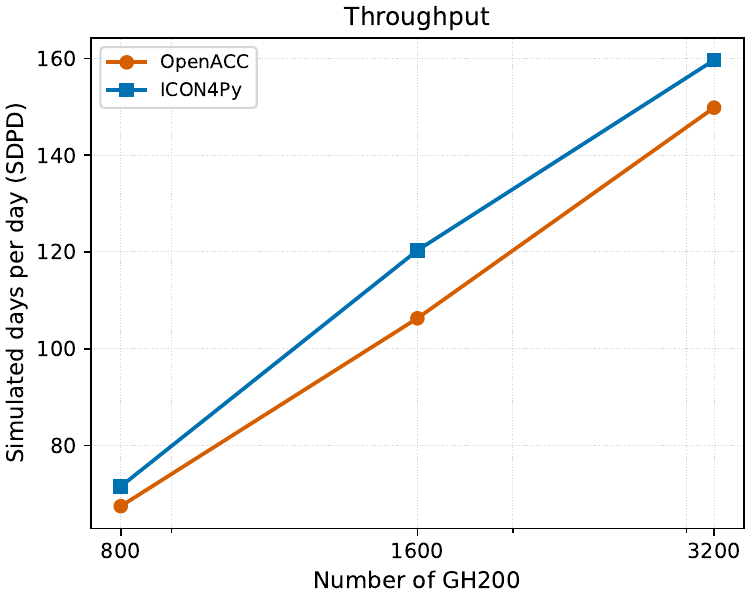}
    \caption{Throughput of the R2B10 atmosphere-land-ocean simulation with an achieved throughput of 160 simulated days per day for the setup with the ICON4Py dynamical core.}
    \Description[SDPD for R2B10 coupled]{Throughput of the R2B10 atmosphere-land-ocean simulation with an achieved throughput of 160 simulated days per day for the setup with the ICON4Py dynamical core.}
    \label{fig:scaling_strong_coupled_r2b10}
\end{figure}

In this Section we describe a coupled atmosphere-land-ocean simulation as representative of a physical climate model. The setup is based on the eXtended Predictions and Projections (XPP) configuration of ICON \cite{Muelleretal2025}, and it is a production quality simulation setup. The XPP configuration includes ICON NWP atmospheric components, coupled to the ICON ocean component \cite{Kornetal2022}, including sea ice, and coupled to the JSBACH land surface model \cite{Schnecketal2022}, including its internal component for hydrological discharge to the ocean. The atmosphere and land surface components run on GPUs, while the ocean and sea ice run on CPUs and coupling is performed by the Yet Another Coupler (YAC) \cite{Hankeetal2016}. Given the numerical characteristics of the simulated systems, the atmosphere uses a time step of 20 seconds whereas the ocean uses a larger time step of 120 seconds. Ocean and Atmosphere are coupled every 600 seconds. Having the Ocean on CPUs is the setup of choice, since the dynamics of the Ocean requires much less computations than the atmosphere. The land model is also not computationally demanding, but the coupling with the atmosphere is much tighter, and hence having it on GPUs (this is done using the OpenACC implementation) keeps the data exchange local to the accelerator. 

Figure \ref{fig:scaling_strong_coupled_r2b10} shows the achieved throughput in simulated days per day for a R2B10 grid (2.5Km global resolution) using 120 vertical levels in the atmosphere and 72 in the Ocean. The ICON4Py implementation of the dynamics makes the overall simulation perform 10\% better than the original Fortran+OpenACC version.

\section{Conclusions}
\label{sec:conclusion}

In this paper we present how we manage to integrate a performance critical Python computation into a traditional Fortran HPC application. The application is the ICON modeling system and the computation is its dynamical core, written in a package called ICON4Py~\cite{icon4py}. We demonstrate that the performance benefit, at the current stage of optimization, is between 20 and 30\% in the dynamical core. In addition to the performance benefit, the advantage of the approach is twofold: on one side the development of the dynamical core has been carried out independently from the main application conventions. This allows for developing components independently and can demonstrate a path to incrementally modernize legacy software. On the other hand, the Python implementation, that lives and can be tested independently, can find other applications in the Python ecosystem.

We believe that this modular approach can be applied to other computing fields in order to shorten the gap from prototyping and deployment, and for opening up integration possibilities that would otherwise be difficult to pursue.

\section{Acknowledgments}
The text in the paper has been reviewed using Google Gemini and Claude LLMs and the plots have been generated by scripts generated by Claude LLM. In both cases, the text and the code have been carefully reviewed and adapted to reflect the intention of the Authors for the text and the correctness of the scripts for the plots.

\printbibliography

\end{document}